\documentclass{article}

\usepackage{arxiv}

\usepackage[T1]{fontenc}
\usepackage[utf8]{inputenc}

\usepackage{csquotes}
\usepackage[english]{babel}

\usepackage{xcolor}
\usepackage{graphicx}
\usepackage{tikz}

\usepackage{bm}
\usepackage{amsmath}
\usepackage{amsthm}
\usepackage{amsfonts}
\usepackage{amssymb}
\usepackage{microtype}

\usepackage{siunitx}

\usepackage{url}
\usepackage[colorlinks=false,hidelinks]{hyperref}

\usepackage[style=ieee]{biblatex}
\usepackage[acronym,hyperfirst=false,nohypertypes={acronym}]{glossaries}
\usepackage{cleveref}

\usepackage{caption}
\usepackage{subcaption}

\usepackage{tabu}
\usepackage{booktabs}

\usepackage[disable]{todonotes}

\graphicspath{{./figures}}

\crefformat{equation}{(#2#1#3)}

\theoremstyle{definition}
\theoremstyle{plain}
\theoremstyle{remark}
\crefname{definition}{Definition}{Definitions}
\crefname{figure}{Figure}{Figures}
\crefname{table}{Table}{Tables}
\crefname{section}{Section}{Sections}

\newacronym[shortplural={CPS}]{CPS}{CPS}{cyber-physical system}
\newacronym{AI}{AI}{Artificial Intelligence}
\newacronym{AL}{AL}{Adversarial Learning}
\newacronym{ANN}{ANN}{Artificial Neural Network}
\newacronym{ARL}{ARL}{Adversarial Resilience Learning}
\newacronym{GAN}{GAN}{Generative Adversarial Network}
\newacronym{GPGPU}{GPGPU}{General-Purpose Graphics Processing Unit}
\newacronym{DL}{DL}{Deep Learning}
\newacronym{GRU}{GRU}{Gated Recurrent Unit}
\newacronym{ICT}{ICT}{Information and Communication Technology}
\newacronym{LSTM}{LSTM}{Long-Short Term Memory}
\newacronym{AEG}{AEG}{Attack Execution Graph}
\newacronym{ML}{ML}{Machine Learning}
\newacronym{PoC}{PoC}{Proof-of-Concept}
\newacronym{RL}{RL}{Reinforcement Learning}
\newacronym{DNN}{DNN}{Deep Neural Network}
\newacronym{RNN}{RNN}{Recurrent Neural Network}
\newacronym{CNN}{CNN}{Convolutional Neural Network}
\newacronym{SIMD}{SIMD}{Single Instruction, Multiple Data}
\newacronym{DoE}{DoE}{design of experiments}
\newacronym{DNC}{DNC}{Differentiable Neural Computer}
\newacronym{PV}{PV}{Photovoltaic}
\newacronym{STL}{STL}{Signal Temporal Logic}
\newacronym{MTL}{MTL}{Metric Temporal Logic}
\newacronym{MITL}{MITL}{Metric Interval Temporal Logic}
\newacronym{AV}{AV}{Autonomous Vehicle}
\newacronym[shortplural={MAS}]{MAS}{MAS}{Multi Agent System}
\newacronym{RTU}{RTU}{Remote Terminal Unit}
\newacronym{TCP}{TCP}{Transmission Control Protocol}
\newacronym{LPEP}{LPEP}{Lightweight Power Exchange Protocol}
\newacronym{FMI}{FMI}{Functional Mock-Up Interface}
\newacronym{XML}{XML}{eXtensible Markup Language}

\newglossaryentry{resilience}{
  name=resilience,
  description={The ability of a system to withstand unforseen, rare and
  potentially catastrophic events and to self-heal or improve in
  reaction to these events. Ideally resilience is increasing monotonously.}
}

\title{Analyzing Cyber-Physical Systems from the Perspective of Artificial
  Intelligence}
\author{%
  Eric MSP Veith,
  Lars Fischer,
  Martin Tröschel\\
  OFFIS e.V.\\
  Power Systems Intelligence\\
  Escherweg 2\\
  26121 Oldenburg, Germany\\
  Email: \texttt{firstname.lastname@offis.de}
  \And
  Astrid Nieße\\
  Leibniz University Hannover\\
  Department of Energy Informatics\\
  Appelstr. 9A\\
  30167 Hannover, Germany\\
  Email: \texttt{niesse@ei.uni-hannover.de}}

\begin{document}
\maketitle

\begin{abstract}

  Principles of modern \gls{CPS} analysis are based on analytical methods that
  depend on whether safety or liveness requirements are considered. Complexity
  is abstracted through different techniques, ranging from stochastic
  modelling to contracts. However, both distributed heuristics and
  \gls{AI}-based approaches as well as the user perspective or unpredictable
  effects, such as accidents or the weather, introduce enough uncertainty to
  warrant reinforcement-learning-based approaches. This paper compares
  traditional approaches in the domain of \gls{CPS} modelling and analysis
  with the \gls{AI} researcher perspective to exploring unknown complex
  systems.

\end{abstract}
\glsresetall

\keywords{Cyber-Physical Systems \and Neural Network Control \and Multi-Agent
Systems \and Reinforcement Learning \and Cyber-Physical Systems security}

\section{Introduction}\label{sec:introduction}

The notion of \glspl{CPS} describes the combination of \gls{ICT} and software
(the ``cyber'' part) with physical components. A \gls{CPS} can emerge from
embedded systems by internetworking them. The first big research program
focusing on \gls{CPS} has been started by the US National Science Foundation
in 2006, where the term \gls{CPS} is defined in as such that it ``refers to
the tight conjoining of and coordination between computational and physical
resources,'' stating ``[w]e envision that the cyber-physical systems of
tomorrow will far exceed those of today in terms of adaptability, autonomy,
efficiency, functionality, reliability, safety, and
usability''~\parencite{NSF10515}.

While the notion of \glspl{CPS} by the U.S.~National Science Foundation, as
outlined above, includes \gls{ICT}, it does not explicitly name \gls{AI} as a
necessary component to raise an embedded system to the status of a \gls{CPS}.
Yet, the availability of sensory data together with a communications system
and the ability to exert actions upon the physical world that have been
planned for the whole compound of embedded systems components readily suggests
that issues of planning, the increase of reflectivity, efficiency, and
lowering resource usage is achieved by increasing the ``intelligence'' of the
overall system. As such, researchers in the domain of \gls{AI} have found
numerous application domains.

However, the two worlds of \gls{CPS} and \gls{AI} usually operate on different
terms: \glspl{CPS} require operation within well-defined boundaries, i.e., as
far as possible deterministic behavior within well-known, strictly enforced
margins of error. In contrast, many \gls{AI} techniques---\glspl{ANN}
foremost---are firmly rooted in the domain of statistics, which is probably
very well seen in the \gls{ANN} training process.

There is already some history in investigating how \gls{AI} can form an
integral part of a high-assurance system, such as \glspl{CPS} typically
are. This is the core of the first perspective we outline in this paper,
namely, the research question of how to bring systems with \gls{AI}-components
whose verification is inherently difficult, such as distributed heuristics
employing \glspl{MAS} or \glspl{ANN}, into the domain of \gls{CPS}, where
verification methods are necessary to avoid high risks in terms of costs, or
even life. Examples from the power system domain
are COHDA~\cite{hinrichs2013cohda} or
Winzent~\cite{veith2013lightweight,ruppert2014evolutionary,veith2017universal,veith2017agent}
that both manage real power generation and consumption schedules using a
distributed approach. In the case of COHDA, the system constitutes a
distributed heuristic, whereas Winzent is deterministic, but relies on
\glspl{ANN} for generation and consumption forecasts.
\textcite{schumann2010applications} compiled further contributions in which
the authors first revisit the robust control, then discussing in all following
chapters topics such as \gls{ANN} complexity analysis; control system design
and test; stability, convergence, and verification; as well as anomaly
detection featuring \glspl{ANN} at their core. The applications range from
automotive, submarine and aircraft control to power systems management and
medical systems. A more recent summary, albeit focusing exclusively on the
domain of highly automated vehicles, is given by \textcite{Damm2018}.

This paper is written by and for \gls{AI} researchers, carefully reviewing the
literature from two perspectives: First, the inclusion of \gls{AI} components
into \glspl{CPS}, outlining challenges and recent contributions, emphasizing
the differences and specific requirements of the \gls{CPS} domain to \gls{AI}.
This first aspect considers the analysis of \gls{AI} techniques such as
\glspl{MAS} or \glspl{DNN} in the context of \gls{CPS}, where the complexity
in inherent statistical non-determinism of \gls{AI} components need to be
tamed.  The second aspect turns the tables on the \gls{CPS}-\gls{AI}
relationship and considers the increase of stability of a \gls{CPS} through
\gls{AI}: Here, \gls{AI} is used to analyze aspects in the behavior of
\glspl{CPS}, reaching points in a complex search space of systems behavior
that are hard to conquer with traditional analysis methods.

In the domain of \gls{CPS}, two key types of requirements are considered:
Liveness and safety requirements. The first is colloquially expressed as
``something good eventually happens,'' whereas the mnemonic for the latter is
``nothing bad ever happens.'' Clearly, considering those requirements, the
focus rests on safety requirements with regards to \gls{AI} in the context of
\glspl{CPS}; we will honor this and touch liveness requirements rather
lightly where they are connected to safety requirements.

The remainder of this paper is structured as follows: We give a brief
introduction to the relevant fundamentals of \gls{CPS} modelling and analysis
in \cref{sec:cps-analysis-primer}. We then describe how programs can be
derived from a formal specification in \cref{sec:introduction}. In
\cref{sec:ann-control-falsification}, we provide a summary how \glspl{ANN}
that serve as the controller in a \gls{CPS} are being tested for safety
requirements. We move on to a whole-system view in \cref{sec:introduction} and
present common simulation frameworks. Extended the topic from \gls{DL} to a
broad \gls{AI} perspective, we take \glspl{MAS} into consideration in
\cref{sec:introduction}, paying specific attention to the communication
between distributed systems. Finally, \cref{sec:introduction} outlines
techniques to derive attach vectors against specific \glspl{CPS} and summarize
recent efforts for automatic analysis of \glspl{CPS} through \gls{AI} methods.

\section{Analyzing Cypber-Physical Systems for Safety Requirements: A Primer}
\label{sec:cps-analysis-primer}

In this section, we give a brief primer on the fundamentals relevant in the
context of the work at hand. It serves mainly as an introduction for scientists from
other domains, most notably \gls{AI}, who are not familiar with the approaches
and basic assumptions of \gls{CPS} analysis. Interested readers are referred
to textbooks such as the one by \textcite{alur2015principles}.

When specifying a \gls{CPS}, one needs to consider liveness and safety
requirements. These two express different types of requirements, memorizable
through the two sentences given in the introduction.
\textcite{alur2015principles} gives a simple example that illustrates the
difference very well. Consider a railway with two tracks, one for each
direction, leading to a bridge. The bridge is narrow such that only one track
fits on it, i.e., the two tracks merge at both ends of the bridge. This
critical section is, of course, guarded by signals on both ends. The
\emph{safety requirement} of this simple system can be expressed by the
following property:

\begin{equation*}
  \mathtt{TrainSafety}:\quad \neg\left[(\mathit{mode}_W = \mathtt{bridge})
    \wedge (\mathit{mode}_E = \mathtt{bridge})\right]~.
\end{equation*}

Hence, two trains arriving at the west end (W) and east end (E) of the bridge
must wait; thus, the safety requirement ensures that ``nothing bad ever
happens.'' However, the property does not present a solution: The two trains
halt, but from the property alone, none enters the bridge. This requirement
that ``one train eventually enters the bridge'' is expressed by the system's
liveness requirements, i.e., ``something good eventually happens.'' Notice the
presence of the word ``eventually,'' which indicates a temporal dimension to
the situation that is sorely missing from the simple property above.

Any \gls{CPS} does not simply come to live in the form of source code or
hardware; instead, it relies on a formal specification first. Formal languages
allow to express the behavior of a system. The formal specification serves as
the basis to reason about the system itself and is also the ultimate tool to
verify whether an actual implementation follows the intended behavior or not.
One of the commonly used specification formalisms is temporal logic,
specifically its variant \gls{MTL}. The \gls{MTL} formalism consists of
propositional variables, the logical operators \(\neg\) and \(\vee\), and the
temporal modal operators \(U_I\) (``\(\varphi\) until in \(I \psi\)'') and
\(S_I\) (``\(\varphi\) since in \(I \psi\). Here, \(\varphi\) and \(\psi\)
represent any valid formula in \gls{MTL}, and \(I\) denotes a temporal
interval. If \(I\) is omitted, \([0; \infty)\) is implicitly assumed.
\gls{MITL} adds ``syntactic sugar'' to \gls{MTL} by replacing commonly used
constructs with dedicated operators. According to
\textcite{ouaknine2005decidability}, \gls{MTL}/\gls{MITL} are the dominating
formalism for describing real-time systems.

The task of falsifying a system is coupled with a \gls{CPS}' safety
requirements, i.e., ``nothing bad ever happens.'' Obviously, if just one
counterexample can be found for a specification, then the system's safety
requirements are defied. Consequently, for many \gls{CPS} that harbour
\gls{ML} components, finding inputs that yield grotesquely wrong outputs is a
required, but not a simple task. Temporal logic has a long history;
\textcite{ohrstrom2007temporal} give a very good account in this regard.

In first-order logic, an expression over a set of variables \(V\) is evaluated
with respect to the valuation for \(V\). In temporal logic, a formula
\(\varphi\) is valuated with respect to an infinite sequence of valuations
over \(V\). The problem of checking whether a given model and its behavior
satisfies its specification expressed in temporal logic by at least one
formula \(\varphi\) is called \emph{model checking}. Model checking has been
developed independently by Clarke and Emerson (the book by
\textcite{clarke2018model} is an integral part of the standard literature
corpus in this field), and
\textcite{queille1982specification,queille1983fairness} in the early 1980s.
When checking for safety requirements, one tries to \emph{falsify}, i.e., find
a valuation for which the model and specification differ. Obviously, only one
such case is enough to defy a \gls{CPS}' safety requirements. Algorithmic
falsification techniques have been implemented, employing stochastic search
strategies and nonlinear optimization methods. These pieces of software, such
as Breach~\cite{donze2010breach}, S-TaLiRo~\cite{annpureddy2011s},
C2E2~\cite{duggirala2015c2e2}, and RRT-REX~\cite{dreossi2015efficient} are
mostly the basis or baseline for novel approaches presented in
\Cref{sec:ann-control-falsification}.

Until now, we have assumed that the specification \emph{and} the model---or
even the system---already exist. Considering green-field projects, this isn't
the case, of course. Then, one usually starts out by sketching the
requirements. An earlier publication by \textcite{clarke1981design} considers
the process of arriving at a program expressed as a flowchart. Finally, to
follow the line of thought further, the specification itself must be correct.
Checking theorems for their correctness is called \emph{theorem proving}.
Employing programs for this, i.e., enabling \emph{automated theory proving},
has a long history---being essentially based on the works by Aristotle,
Frege~\cite{zalta1998frege,frege1950grundlagen,wille2018gottlob},
\textcite{russell1978principia}, and Skolem and
Löwenhein~\cite{van1967frege,badesa2009birth}---and has made enormous advances
in the past years, being now routinely employed in the industry. E.g., the
Intel Pentium FDIV bug~\cite{cipra1995number} has firmly established automated
theorem proving in the CPU industry. Examples for concrete implementations
come from \textcite{gordon1993introduction}, utilizing higher-order logic, and
\textcite{owre1992pvs}. An in-depth overview and examples are offered by
\textcite{kaufmann2013computer}.

In checking all stages of the design process, from creating the model,
deriving the implementation, checking the implementation's requirements and
finally monitoring its behavior during runtime, one tries to achieve
\emph{end-to-end correctness} of the \gls{CPS}. There is no fixed recipe for
all steps; \textcite{B2017} present a not too old approach from
traditional \gls{CPS} design where \gls{ML} does explicitly not form an
integral part of the \gls{CPS}, which is why it is exactly a good example in
this regard. The example the authors present is a surveillance drone that
patrols a fixed route via waypoints. In their approach, the (reactive)
robotics software is first tested via model checking. Then, the safety
requirements are noted in \gls{STL} and verified to hold continuously at
runtime.

Another factor that makes the previously mentioned publication a good example
is the programming language the authors employ for the software: \emph{P}.
Presented by \textcite{Desai}, its main paradigms are asynchronous control
flow that is event-driven. As textbooks about \glspl{CPS} usually explain in
the very first chapters, those systems are \emph{reactive}---probably one of
the major paradigm gaps between the two domains, \gls{CPS} and \gls{AI}, since
a major part of the latter strives to develop \emph{proactive} systems.

\textcite{Seshia2016} bring this difference to the point, emphasizing that the
concept of \emph{verified \gls{AI}} that would fit the \gls{CPS} domain would
have ``strong, ideally provable, assurances of correctness with respect to
mathematically-specified requirements,'' while \gls{ML} methods and models
such as \glspl{DNN} that are being trained with ``millions of data points''
can exhibit close to stochastic behavior. The authors argue that bridging this
gap can be done through training and test data generation. The focus on data
and how it interacts with \gls{ML} models is further deepened in
\cref{sec:introduction}.

\section{Synthesis Methods}
\label{sec:synthesis-methods}

Considering the previous section, it becomes obvious that reasoning about a
\gls{CPS} involves a formal specification as well as a program that follows
it. The specification is a high-level description of what the final system
should do and ideally, the implementation in hard- and software follows this
specification. Bridging these two levels of abstraction poses not one, but two
questions: First, does a piece of software that was written according to a
specification match the intended behavior expressed by this specification in
all cases? And second, is there instead a way to generate a program that
complies with the high-level specification? Formal synthesis is the latter
process, i.e., generating a program from a high-level specification.

\textcite{Jha2017} present a framework in this regard that combines \gls{ML}
and formal synthesis, i.e., a framework in inductive synthesis. Inductive
synthesis---described, e.g., in by \textcite{Gold1967},
\textcite{shapiro1982algorithmic}, and
\textcite{summers1977methodology}---seeks to find a program that matches a set
of input/output pairs. At high level, inductive synthesis is an instance of
learning from examples, commonly known also as inductive inference or machine
learning. This connection can be deduced very easily from the works of
\textcite{angluin1983inductive}, and \textcite{russell2016artificial}. The
framework proposed treats the synthesis as a problem of language learning.
However, \textcite{Jha2017} also clearly note the differences between a
program synthesis and the usual \gls{ML} approach, which manifests itself not
just in concept classes the learning algorithms are applied to, but
specifically in the difference of \emph{exact learning} and \emph{approximate
learning}.  The latter is the typical \gls{ML} task, where driving the
training error down to zero is often not necessary or even not desirable (as
it would indicate overfitting), whereas a program synthesis cannot get away
with being a correct program 98\% of the time.

Another aspect of synthesis methods is parameter synthesis.  In this case, a
formal description exists, e.g., in \gls{STL}, as well as a model, but the
\gls{STL} formulæ lack concrete signal or time values. This means that the
general behavior of a system is known, but threshold or critical values are
not. \textcite{Jin2015} propose an algorithm to mine requirements from
closed-loop control models, in which the inputs to their algorithm is the
plant model as well as a requirement template expressed in parametric
\gls{STL}. The algorithm guarantees a tight formulation of the parameters,
i.e., parameter values are as close to the actual limits of the system as
possible. The overall approach is helpful to, e.g., validate future versions
of the \gls{CPS}. 

\section{Neural Control Falsification}
\label{sec:ann-control-falsification}

\emph{Neural control} describes a \gls{CPS} in which an \gls{ANN} takes the
role of a controller, replacing a complex traditional controller, such as a
closed-loop system. Their falsification is based on the system's safety
requirements, i.e., the part expressing that ``nothing bad every happens.''
Consequently, one counterexample defies this, which is ultimately the goal of
any falsification technique. Whatever ``bad'' means is dependent on the actual
system and the world it is being deployed in; in an \gls{AV}, this can range
from the misclassification of an object or human being to an actual collision.

When an \gls{ANN} is used as a controller, problems of \gls{AL}, as they are
studied in the domain of \gls{AI}, of course also apply to the use \glspl{ANN}
as controllers in \glspl{CPS}. In \gls{AL}, an adversarial sample constitutes
a sample that is presented to an \gls{ML} and that features only minimal
perturbation compared to a legitimate sample, but causes the \gls{ML} model to
exhibit a starkly wrong result. Well-known examples of \gls{AL} have been
observed in e-mails in the early days of spam filtering, where a blacklisted
word such as \emph{viagra} was modified to read \emph{víâg®á} and therefore
still readable to humans as the same word, but caused to offending e-mail to
escape spam filters. The problem emerges especially in the domain of deep
learning, where complex \glspl{ANN} can, e.g., misclassify images when only a
minimal amount of RGB noise is added to the picture. 

The foundation of computer vision that is prominently used for \glspl{AV} are
\glspl{CNN}. Although they have a long history and automatic training with
backpropagation of error has been shown by \textcite{LeCun1989} in 1989, only
the advent of \glspl{GPGPU} have made effective training possible. The
well-documented work by \textcite{cirecsan2010deep} marks one of the major
achievements in this regard. However, effective adversarial samples have been
quickly found; an impressive account has been made by \textcite{Nguyen2015}.
\textcite{Goodfellow2014} offer an in-depth explanation of \gls{AL} in the
context of \gls{DL}, arguing that the primary reason for the effectiveness of
adversarial samples is the linear nature of certain \glspl{ANN}.

Recently, researchers have worked on methods to test \glspl{ANN} against
adversarial examples, both with and without knowledge of the neural network's
inner structure. \textcite{Papernot2017} thoroughly demonstrate practical
black-box attacks after previously noting the general limitations of \gls{DL}
in adversarial settings~\cite{Papernot2016}. \textcite{Chen2017} recount how
black-box attacks use substitute models trained on the target \gls{ANN},
exploiting the fact that adversarial samples (e.g., adversarial images) are
highly transferable. In their paper, the authors then employ zeroth order
optimization methods to directly estimate the gradients of the target model to
effectively generate adversarial samples without needing a substitute model.

Where black-box testing of \glspl{ANN} assumes no prior knowledge of the
networks themselves, white-box tests specifically trace the activation of
neurons given certain inputs in order to derive minimal changes to valid
samples to arrive at adversarial samples. With \emph{DeepXplore}, one of the
first frameworks for white-box testing of \gls{DL} \glspl{ANN},
\textcite{Pei2017} present a framework that introduces neuron coverage to
measure the activation of parts of the \gls{ANN} in order to then derive
adversarial samples. Manual testing of correct or incorrect behavior of the
overall system is achieved by cross-linking oracles, which, in this case, are
other \gls{DL} systems with similar functionality. In this work, deriving
adversarial samples with high neuron coverage is represented as optimization
problem that is subject to efficient gradient-descent methods.
\emph{DeepXplore} has already been the basis for several \gls{CPS}-specific
frameworks, such as \emph{DeepTest} by \textcite{Tian2017} that aims at
testing \glspl{AV} driven by \glspl{DNN}.

With \emph{AI\textsuperscript{2}}, \textcite{Gehr2018} propose a sound
analyzer for \glspl{DNN}. The key building block is the transformation of
operations in the \gls{ANN} to conditional affine transformations, i.e.,
affine transformations guarded by logical constraints. This allows to treat
the elements of the \gls{DNN} in abstract domains such that property
verification, specifically of robustness properties, is possible. As reasoning
of robustness and safety properties of \glspl{ANN} is enabled through
AI\textsuperscript{2}, this bridges the world of \glspl{ANN} and classic
abstract reasoning of \glspl{CPS}. One side effect of the proposed methodology
is the restriction of \emph{AI\textsuperscript{2}} to only certain kinds of
activation functions and network layouts, namely ReLU, max pooling, and
convolutional layers that make a feed-forward/convolutional neural network.

All white-box falsification methods currently attack feed-forward
\glspl{ANN}, with \glspl{CNN} being a variant thereof. This makes sense,
considering that \glspl{CNN} find application in the computer vision
discipline, where \glspl{CNN} are currently the best structure for
image/object classification. With Capsule Networks that differ from
\glspl{CNN} especially through their dynamic routing concept between capsules,
\textcite{Sabour2017} have proposed a \gls{ANN} architecture that addresses
some of the shortcomings of \glspl{CNN}, e.g., that a \gls{CNN} happily
detects a normal human face even if the eyes are located on the chin. However,
capsule networks still cannot be efficiently trained; but of course, this is
being worked on~\cite{xi2017capsule,Wang2018}. Another reason for tackling
feed-forward networks first is the inherent complexity of other \gls{ANN}
architectures: Feed-forward networks approximate any Borel-measurable
function; for this claim exist no less than three proofs, the most often cited
ones being developed by \textcite{hornik1989multilayer}, and
\textcite{cybenko1989approximation}.

In between white- and black-box testing is gray-box testing, in which some
knowledge of the underlying model is assumed to be known, but no full
analytical coverage is desired. \textcite{Dreossi2017} develop a hybrid,
gray-box approach, assuming some internal knowledge of the \gls{ML} components
within the \gls{CPS}. The authors assume these components to be classifiers,
specifically binary classifiers, rightfully implying that any multi-class
classifier can be turned into a binary classifier without loss of generality.
The two parts---the \gls{CPS} specified in \gls{STL}, and the \gls{ML}
component---are then separately analyzed. Then, they assume a variant of the
\gls{CPS} model \(M\) the authors denote with \(M^{+}\), in which the \gls{ML}
component is perfect. From this point, considering only the \gls{CPS}
component described by the \gls{STL} formula \(\varphi\), the two sets of
inputs to \(M^{+}\) \(U^{+}_{\varphi}\) and \(U^{+}_{\neg\varphi}\) are
computed. The second part of the falsification engine then tries to identify
inputs contained in \(U^{+}_{\varphi}\) for which the \gls{ML} component
fails. The \gls{ML} classifier is replaced by a simpler approximation
operating on an abstracted feature space; sampling of the desired points from
the thus simplified feature space is done with quasi-Monte Carlo techniques
building on the discrepancy notion from equidistribution
theory~\cite{weyl1916gleichverteilung,rosenblatt1995pointwise}. The authors'
case study utilizes the Unity-Udacity \gls{AV} simulator in which a \gls{CNN}
needs to detect a cow and initiate emergency braking.

In a similar vein, \textcite{Yaghoubi2019} present a state-of-the-art,
effective framework in this regard. Additionally to the black-box approach of
knowing inputs and outputs, they extract dynamic model linearizations along
the systems' trajectories, arguing that this kind of information is readily
available via, e.g., the Simulink linear analysis toolbox anyways. This
linearization around the trajectories is used to apply gradient descent in
order to yield a valid input similar to the original one, but with a negative
robustness value, i.e., the falsification. The input search space is not
constraint by this approach, which is a further advantage.  The authors claim
that their framework outperforms black-box system testing methods, showing in
case studies shorter times to falsification as well as consistently finding
falsifications where the methods chosen for comparison fail. In their
experimental results, they choose uniform random sampling and simulated
annealing implementations of S-TaLiRo~\cite{annpureddy2011s} the baseline to
beat. Sadly, black-box testing methods such as the previously discussed one by
\textcite{Chen2017}, a comparison with approach by
\textcite{Dreossi2017}---which the authors explicitly cite---, or even
white-box testing methods, are absent from the case studies.

The most striking difference between the two frameworks by
\textcite{Dreossi2017}, and \textcite{Yaghoubi2019} lies in utilizing ideas
from optimal control
theory~\cite{abbas2014functional,zutshi2014multiple,Li2017,Yaghoubi2019},
meaning that the frameworks consider system-level building blocks instead of
only the controlling \gls{ANN}. An advantage of this approach in comparison to
\emph{DeepXplore}, \emph{AI\textsuperscript{2}}, and similar white-box
methods, however, is their framework's ability to falsify \gls{RNN}
controllers, which \textcite{Yaghoubi2019} also show in one case study;
\textcite{Dreossi2017} specifically refer to \emph{DeepXplore} with regards to
this fact.

Future directions seem already clearly dictated.
\textcite{seidl1991dynamicsystem,siegelmann1995computational} have shown that
\glspl{RNN} are approximators of dynamic systems, a fact accommodated by
\textcite{Yaghoubi2019}. Apart from serving as controllers, applications for
\glspl{RNN} are twofold: One is the classical forecasting from historical
data. One such forecasting that can serve as the example for the whole concept
is short-term load forecasting in power grids, where the decisions derived
from the forecast of an \gls{RNN} obviously influence the behavior of the
\gls{CPS} power grid.  \textcite{bianchi2017recurrent} give a recent
comparison of Elman simple \glspl{RNN}~\cite{elman1990finding}, Wavelet
networks~\cite{cho2014learning}, \gls{LSTM} cells~\cite{hochreiter1997long},
and \glspl{GRU}~\cite{cho2014learning}. The second aspect calls on the notion
of the \emph{digital twin}, i.e., ``a dynamic virtual representation of a
physical object or system across its lifecycle, using real-time data to enable
understanding, learning and reasoning,'' according to
\textcite{bolton2018customer}. This idea specifically enables Deep \gls{RL},
which has become a huge topic of interest since the hallmark publication of
\textcite{mnih2013playing}. Clearly, the ``trial and error'' approach of
\gls{RL} calls for a simulation model in the context of critical \glspl{CPS}.
Finally, the \gls{DNC}, proposed by \textcite{graves2016hybrid}, with an
\gls{RNN} serving as a controller in a differentiable Von-Neumann architecture
with external memory, promises to approximate algorithms, further widening the
capabilities of \gls{DL} modules that will find their way into \glspl{CPS}.

\section{Simulation-based Testing Frameworks}
\label{sec:simulation-frameworks}

Once the environment becomes increasingly complex, analytical methods as
mentioned in the previous sections fall short in providing real-world, i.e.,
realistic inputs. While the methods mentioned in
\cref{sec:ann-control-falsification} can be used to derive inputs that falsify
a model, they are not necessarily guaranteed to actually appear in a
production environment. Consider an autonomous vehicle as the \gls{CPS}:
Another car flying upside down from one building to another might certainly
trigger a wrong driving decision by the neural controller, but are not very
likely to be perceived in a real-world environment.

In this regard, \textcite{Tuncali2018} propose simulation-based generation of
adversarial samples. In contrast to traditional synthesis or verification
methods, in which an understanding of the system exists and is expressed by,
e.g., \gls{STL} or contracts, simulation-based analysis treats the \gls{CPS}
as black-box model. The authors describe methods to pertubate testing scenario
parameter configurations, i.e., they seek to find scenarios that lead to
unexpected behavior. Their research is motivated by the status quo of
\gls{ANN} verification that is only very limitedly possible as \glspl{ANN}
correspond to complex nonlinear and non-convex functions. Even though the
generation of adversarial examples is studied intensively, the authors explain
that simulation-based generation of adversarial samples constitutes a similar
problem in the name only: \Gls{AL}, as it is studied in the domain of deep
learning, focuses on one \gls{ANN} and therefore deals with the falsification
of an \gls{ANN} at component level, whereas the proposed simulation-based
approach is a falsification approach at system level, where the \gls{ANN} is
just one component in a complex \gls{CPS}.

\textcite{Kelly2017} argue that a simulation might work with synthetic data or
variations on the simulation scenario configuration, but that this still
leaves an enormous search space in which realistic situations represent only a
smaller portion of the overall data. They add that collisions or blandly wrong
decisions by a neural network controller are interesting, but far more often
dangerous or ``near-miss'' situations in an \gls{AV} driving scenario are just
as interesting, but harder to detect. The authors acknowledge that
sophisticated world simulators are necessary to provide the \gls{AV}'s
perception pipeline with realistic data and, based on this, incorporate the
video game \emph{Grand Theft Auto V} into their simulator as the world in
which the \gls{AV} is driving. Their approach also differs in that the
simulator does not explicitly force dangerous situations, but instead tries to
log sufficient miles for an expressive test.

In general, pertinent literature stresses that not the software, but the model
of the simulated world is the major challenge in any simulation task.
\textcite{Lobao1997}, and Robinson~\cite{robinson1994simulation,Robinson2004}
specifically argue how, as scope and complexity of the model increase, model
confidence and even accuracy finally decrease. \glspl{CPS}, with a combination
of different domains, are also prone to suffer from another problem with
regards to simulation: There is no single simulation software that covers
multiple domains. An \gls{ICT} simulator cannot cover the intricacies of
simulating a power system or present the perception pipeline of an \gls{AV}
with realistic images. To address this problem, co-simulation is a viable
approach. It assumes that the best possible course is to couple
domain-specific simulators, synchronizing them and allowing them to exchange
data. \textcite{gomes2017co} offer a more detailed argumentation and
description of the specific challenges;
\textcite{nguyen2017conceptual,palensky2017cosimulation} focus on the
specifics of co-simulation in the power systems domain.

With regards to software that facilitates co-simulation, radically different
approaches exist. On the simpler end of the spectrum, we can note
\emph{mosaik}~\cite{rohjans2013mosaik}. It synchronizes simulators on a
request-response communication schema, acting as a ``data kraken'' in which
the experimenter connects communicating models from different simulators via
the attributes they announce to \emph{mosaik}; requesting data on a time step
using the \texttt{get\_data} method and delivering data prior to executing a
time step using the \texttt{step} method. mosaik is, therefore, agnostic to
the inner workings of simulators or their models. In contrast,
\emph{Ptolemy~II}~\cite{ptolemaeus2014system} focuses not just on simulation,
but also on modeling and design of concurrent components. Including techniques
such as actor-driven execution, \emph{Ptolemy~II} is a grey-box co-simulation
approach where mosaik treats all simulators as black boxes, requesting only
compliance to the communication protocol as the smallest common denominator.

Besides concrete implementations, co-simulation can be governed by a standard
for support both model exchange and co-simulation of dynamic models. The
\gls{FMI}~\cite{blochwitz2012functional} uses a combination of \gls{XML}
documents and generated C~code to bridge different simulators without forcing
a particular toolset or execution scheme. A good illustration how extensive
co-simulation can be employed in the context of \glspl{AV} as \gls{CPS} is
provided by the key findings document of the ENABLE-S3 EU~research
project~\cite{Leitner2019}.

\section{Multi Agent Systems and Deterministic Communication in Cyber-Physical
  Systems}
\label{sec:mas}

The perspective of \gls{AI} has, until now, been very much focused on
\gls{ML}, giving much space to \gls{DL} models. Obviously, this is too narrow:
\textcite{russell2016artificial} describe \gls{AI} in terms of rational
agents, or, more broadly, ``intelligent entities'' that learn, solve problems,
and act based on their own goals and model of their environment. This is,
considering standard works such as the textbook by
\textcite{wooldridge2009introduction}, where they ultimately distinguish
themselves from the classical \gls{CPS} software: While the latter is always
\emph{reactive}, agents are, except for the simplest forms, \emph{proactive},
meaning that they act, with a specific level of autonomy, towards a goal or to
maximize a utility function regardless of whether they receive input from
their environment or not.  Together with the uncertainty of \gls{ML} models,
this proactiveness provides for the next stark contrast to the traditional
\gls{CPS} domain. Additionally, many agent designs introduce a social
component: There is not one, but several agents, collaborating through a
communication protocol, evolving it to a \gls{MAS}. From the perspective of a
\gls{CPS}, a better---and more specific---name is that of \emph{agent-based
control systems}, combining both aspects, that of autonomy and that of
exerting control over physical components~\cite{bussmann2013multiagent}.

While the basic assumptions of the different domains seem to contradict each
other, the \emph{Divide-et-Impera} approach computer science inheres in
appeals to tackle complex problems even in critical infrastructures and
\glspl{CPS}. Examples include the aforementioned \emph{Universal Smart Grid
Agent}~\cite{veith2017universal}, designs by
\textcite{niesse2012market,Schwerdfeger2014}, where ancillary services,
including frequency control, are provided for power grids through \glspl{MAS},
as well as an earlier summary by \textcite{McArthur2007,McArthur2007a},
crossing to other domains such as autonomous driving, where
\textcite{dresner2005multiagent} employ a \glspl{MAS} for traffic management
at intersections, or \textcite{halle2005collaborative} present a collaborative
driving system modelled through a \gls{MAS}.
When we assume that the previous paragraphs cover the internals of each agent,
what defines the overall behavior of the \gls{MAS} is their communication,
being the agents' \emph{protocol} from the sense of data encoding as well as a
behavioral protocol. \Gls{ICT} protocols are usually far from being
deterministic; the analysis of \textcite{veres2000chaotic} in terms of the
stochastic nature of the congestion control algorithm of the well-known
\gls{TCP} may serve as a pars-pro-toto example.

In general, the \emph{divide-et-impera} approach of \gls{MAS} lets the
foundation of distributed computing surface: \emph{consensus problems}. This
is not solely related to agents with their proactive behavior based on goals
or utility functions; problems such as distributed timekeeping and distributed
snapshotting with classical publications by \textcite{Chandy1985,Mattern} are
well known. In this understanding, software agents constitute the technical
realization of distributed algorithms, paving the way for a more integrated
view on distributed systems and AI. Reaching consensus---i.e., the question of
convergence of a protocol---becomes more difficult to reason compared to
locally running algorithms due to the distributed nature of an \gls{MAS},
where time delays in transmission are not deterministic.
\textcite{Olfati-saber2007} provide a valuable review of newer consensus and
cooperation protocols as well as ways to model consensus as control loops and
using graph theory to qualitatively and quantitatively determine convergence.
Eventual convergence is obviously not guaranteed;
\textcite{hanachi2004protocol,matt2006distributed} show the non-trivial nature
of convergence. Convergence is still different from making sure that a deal is
always possible, as \textcite{faratin1998negotiation} state.

One of the ancestral behavioral protocols for \glspl{MAS} is the \emph{Contract
Net Protcol} by \textcite{smith1980contract}. Here, agents announce tasks using
broadcast messages for other agents to bid on. The announcement also contains
the ranking process, i.e., bids delivered by other agents are ranked according
to metrics such as estimated time to task completion. The announcer, or task
manager, then awards the task to a specific node, informing all other nodes in
the process. The awarded node can then additionally choose to break the task
up into smaller subtasks and sub-contract them through a similar procedure.

The general broadcast-bidding-awarding structure of behavior laid down in the
\emph{Contract Net Procol} has influenced many (negotiation) protocols for
distributed computation. In many cases, additional ideas are brought in to add
efficiency, to speed up the negotiation, or to reduce the amount of messages
or data being sent. The
\gls{LPEP}~\cite{veith2013lightweight,veith2017universal} specifies initial
messages (requests for or offers of power) as broadcasts, but models the
overlay networks the agents use on the power grid in which the agents'
physical entities represent, imposing rules on message routing that limit
message propagation, introducing the concept of dynamic neighborhoods where
supply and demand have as little physical line meter between them as possible,
reducing the line loss. Responses are routed directly through a dynamic
routing table on each node that is being built during the request stage.

Additionally, \textcite{Shen1995} worked towards eschewing the initial
broadcast stage. They employ multicasting---i.e., the network protocol
concept~\cite{rfc1112}---for the task announcement messages, creating interest
groups to which agents can subscribe. \textcite{Wanyama2007} reduce the
number of negotiation rounds until consensus is reached, limiting the scope of
agent coalitions to a \emph{group-choice problem} and basing their negotiation
approach on game theory, replacing explicit knowledge through message
exchanges by implicit knowledge coming from a game-theoretic model of the
negotiation process. \textcite{Garcia2017} have reduced the number of messages
per negotiation, assuming a control theory problem behind the agents'
communication and implementing an asynchronous, event-based protocol based on
a discretized model that is decoupled from the state of the agent's neighbors.

The aforementioned publication by \textcite{Olfati-saber2007} also emphasizes
the effectiveness of neighborhood concepts, based on \emph{small-world
networks} by \textcite{watts1998collective}---being one of the hallmark works
on overlay topologies for distributed computing---, and referring to the
weightings introduced by \textcite{xiao2004fast}. The two works heavily
influenced the later, much-celebrated \emph{small-world model} for \glspl{MAS}
by \textcite{olfati2005ultrafast}. The COHDA protocol by
\textcite{hinrichs2013cohda} builds on the small-world model;
\textcite{niesse2017local} also note that fast convergence or the quantitative
guarantee of convergence do not necessarily mean that the optimal solution to
a problem is found, but that the \gls{ICT} overlay network topology influences
the search for a solution with certain \gls{MAS} protocols.

In the context of a \gls{CPS}, fully decentralized \gls{MAS} approaches to a
problem can be viewed with suspicion. After all, there is no way to control or
``look into'' the process as it happens. The statement of the convergence
problem by \textcite{hanachi2004protocol} mentioned above is approached by the
authors through a \emph{protocol moderator}, i.e., an explicit middleman.
Similarly, for COHDA, \textcite{niesse2016controlled} propose an
observer-controller architecture for the in its core completely
decentralized protocol. The questions these approaches rise is whether how
certain behavior can be formulated as being expected, rather than just
exhibited. It is expressed in the move from specifications to \emph{contracts}
in component design.

The idea of contracts has been and applied by \textcite{Meyer1992} in his
famous and influencing \emph{Eiffel} programming language, where a
\emph{class}---in the sense of object-oriented programming---is considered an
invariant and a class' methods are extended with \emph{preconditions} and
\emph{postconditions} state predicates. As long as the \emph{preconditions}
hold, a method's contracts with the outside world are fulfilled; when the
method's \emph{postconditions} hold, the method fulfills the contracts the
rest of the world expects from it. As such, a \emph{contract} is a component
model specifying what the component expects from its environment, and the
ensuing promises guaranteed by the component under correct use. The ideas
implemented in \emph{Eiffel} root in seminal work by
\textcite{dijkstra1978guarded,lamport1990win} (\emph{weakest preconditions}
and \emph{predicate transformers}); the books by
\textcite{back2000contracts,back2012refinement} are probably the standard
works for reasoning about discrete, untimed process behavior: The
\emph{refinement calculus} describes processes operating on shared variables
using guarded commands.

\textcite{dill1989trace} describes a model based on behaviors exhibited by a
component, called \emph{traces}, that is asynchronous in nature. It has been
extended to a discrete synchronous model by \textcite{dill1994hierarchical};
\textcite{de2001interface} have proposed a concept similar to synchronous
trace structures, called \emph{interface automata}, which has later seen
extension to resources and asynchronous behaviors.  \emph{Process spaces},
developed by \textcite{negulescu2000process}, and the \emph{agent algebra} by
\textcite{passerone2004semantic} offer a more general approach to generic
behaviors and draw heavily from the algebraic approach by
\textcite{burch2001overcoming}. Note that the term ``agent'' in the
\emph{agent algebra} is, ``[\ldots] a generic term that includes software
processes, hardware circuits and physical components, and abstractions
thereof,''~\cite{passerone2004semantic} i.e., not necessarily an autonomous
software agent in line with the descriptions of the previous paragraphs.
Nevertheless, the extensive theory of the \emph{agent algebra} is valuable in
this regard.

Obviously, the notion of ``expectations from the environment'' and ensuing
``guarantees to the environment, given the expectations are met'' are
extremely valuable to reason about components, specifically distributed
components, such as a \gls{MAS} provides. Generally, especially since agents
can form coalitions (i.e., aggregate and disaggregate dynamically), contracts
based on such dynamic components are hard to formulate.
\textcite{Vokrinek2007} propose a protocol that includes checking for contract
compliance and evaluating the overall performance of the contractor. They
include the concept of \emph{penalties} for contractors into the protocol,
making penalties dynamic as part of the bidding process. Realizing that
executing a contract cannot be enforced without implying a specific
reason---the node may become offline due to a failure as well as being
malicious---the authors do not specify what penalties are, or what the result
of a penalty is. I.e., there is no ``penalty announcement'' to other nodes
that would affect further biddings between the contractor and a different
contractee. However, developing such a social dynamic in \glspl{MAS} with all
their implications is no simple task; specifically since one must realize that
this are \emph{ex post} mechanisms, i.e., the damage is already done when
these mechanisms are set in motion.
\textcite{huynh2006integrated,schmidt2007fuzzy} propose trust models for
\glspl{MAS} that try to take initial mistrust into account. However, this may
not be enough for not violating any safety requirement; additionally, a
hitherto fully trustworthy node may fail even though it has a high trust score
(e.g., due to an outage). 

\section{Deriving Attack Vectors on Cyber-Physical Systems}
\label{sec:cps-attack-vectors}

Even when a \gls{CPS} is hardened against adversarial inputs, this does not
necessarily mean that an actual attack is considered. Remember that adversarial
inputs are, in general, any input that foils the model into emitting a wrong
response, such as a misclassification of an image. Adversarial inputs are not
per se malicious, but can constitute such input or a modified input that
triggers false behavior. E.g., \textcite{7498955} analyze the effects of image
quality on \glspl{DNN}; but JPEG compression artifacts do not necessarily
constitute an attack on the model.

In a twist of the \gls{AL} concept, attacks can be staged against the
communication, the very nerves of a \gls{CPS}. This is not only limited to
exploiting bugs in the implementation of protocols, such as parsers; testing
of this lower level of an agent's communication stack can be done using
fuzzers---the work by \textcite{gorbunov2010autofuzz} can be seen as exemplary
of this---, but extends to crafting datagrams that are valid, but expose logic
errors or other unwanted behavior in an \gls{MAS}. Further, the difference
between bad data injection and \gls{AL} in general lies in the target:
\gls{AL} specifically targets an \gls{ANN} serving as controller, taking the
idiosyncrasies of the \gls{ML} aspect into account. Bad data injection
considers the whole \gls{CPS}, trying to fool traditional status monitoring
mechanisms that usually serve as the most trusted input to \gls{CPS}'
controller. For this to work, the exact domain of the \gls{CPS} must be known;
in contrast to the \gls{AL} aspect, that targets the \gls{ANN} controller, but
can otherwise remain agnostic to the domain-specific parts of the actual
\gls{CPS}, bad data injections works because it targets the exact behavior and
inner workings of the monitoring devices. Works that study this kind of
malicious attack are largely based on analyses of undetectable errors---and
not attacks---in otherwise legitimate \gls{CPS} states, such as those
\textcite{clements1981power,wu1989detection} discuss. One of the basic
findings in this regard, shown, e.g., by \textcite{sandberg2010security}, is
that an attacker has to compromise more readings than just the one she
targets.

Specifically, \textcite{Teixeira2010} analyze bad data injection in the
context of cyber security of power grid state estimators. The specific goal is
to inject such data that the state estimation algorithm still converges, i.e.,
the data must appear to be meaningful---specifics on state estimation
algorithms and their convergence behavior can be found in textbooks such as
the one by \textcite{abur2004power}, as well as publications targeting the
apparent convergence problems such as the neural state estimation approach by
\textcite{manitsas2012distribution}---, and thus the attack remains stealthy.
A successful attack diverges the thus assumed state awareness from the actual
state of the power grid. In contrast to an earlier publication by
\textcite{liu2011false}, the authors can relax the attacker knowledge required
to only partial or even outdated knowledge about the power grid, which can be
obtained from the data some \glspl{RTU} send to the control center using
statistical models.  \textcite{Teixeira2010} give an analytical margin on the
amount of knowledge the attacker needs for a successful injection.

How attacks on a \gls{CPS} need only partial information about the \gls{CPS}
itself is demonstrated by \textcite{Ju2018b}. Here, the authors assume that an
attacker has gained control over a generator, such as a wind turbine or
\gls{PV} plant. The attacker's goal is to inflict damage by leveraging the
reactive power control mechanisms present in any medium voltage grid. To
this end, she only needs the general knowledge of the control rules for
reactive power injection or consumption, which is very easy to come by. The
attack works with distributed generators and explicitly without knowledge of
the power grid's layout or other assets present in the grid. Specifically, she
also does not need to compromise the communication between nodes, an aspect
the other works referenced in the previous paragraphs focus on as premise for
the attack to work. The authors show analytically that damage through voltage
disruption based on twice the available reactive power available can be
inflicted by the attacker with nearly no knowledge.

\textcite{Gao2015a} also provide an approach for automated vulnerability
analysis of state estimators. This one, too, is based on false data injection
and considers stealthy attacks. It differs from the investigation by
\textcite{Teixeira2010} through the approach, as it encodes the problem as
logic-based vulnerability analysis, presenting an approach called
\emph{symbolic propagation} to allow successful attacks from only a localized
set of nodes. The authors use satisfiability modulo theory solvers to attack
the thusly formulated problem.

Interestingly enough, these two publications---or others in the same vein---,
although they assume traditional state estimators, are a variant on fooling
controllers through the input data presented to them. Recall that \glspl{ANN}
can approximate functions and \glspl{RNN} dynamic systems, approaches to
replace traditional state estimators with \glspl{ANN} are obvious, with one of
the more recent examples coming from \textcite{manitsas2012distribution}.
These approaches are susceptible to their very own form of \gls{AL}, i.e.,
also suffer from bad data injection.  \textcite{hu2018state} offer a recent
description of this problem.  Countermeasures are being taken, ironically, not
by hardening the \gls{ANN} themselves, as the authors of the publications
discussed in the previous \cref{sec:introduction} have tried to do, but by
using other \glspl{ANN} to detect the attack. \textcite{he2017real} promise
real-time detection of false data injection using a \gls{DNN};
\textcite{mousavian2013real,abbaspour2016detection} undertake a similar
endeavor, differing only in the type of network and data encoding they use.

\glspl{CPS} are not purely technical, but almost always have an economical
perspective. For the power grid, markets and pricing schemes exist that can
subject to exploitation. In this perspective, the attack is not motivated by
the goal to physically destroy a piece of infrastructure, but by maximizing
profits. Not necessarily is this entailed by a mere violation of the rules;
market design is complex enough that a businessman can stage an exploit by
following the rules \emph{by the letter} instead of \emph{by intent}. In this
regard, \textcite{Hirth2018} show how a zonal redispatch market---such as the
model Germany employs---can be gamed. In this market model based on zones
(e.g., north and south), congestion can lead to redispatch of generation or
consumption. If, by the merit order principle, a power plant in the
\emph{north} zone would normally supply power, but cannot, because the link
between the \emph{north} and \emph{south} zones is congested (i.e., the
physical line would be loaded beyond the physical safety limit), the
generation is redispatched to the south, meaning that a more expensive power
plant in the \emph{south} zone supplies the power needed and the redispatched
power plant in the \emph{north} zone receives a compensation for foregone
profits. Using \gls{AI}-based prediction, a power plant operator can predict
redispatch beforehand and enter the initial bidding phase with much lower
variable costs, betting on first awarded the contract to generate money while
also predicting the redispatch to receive the compensation instead of
generating at a loss. This specific form of economic attack is called
\emph{inc-dec gaming}. Similar observations have been made for Great Britain
by \textcite{Konstantinidis2015}.

\textcite{Ju2018b} have shown that very little knowledge of the power grid as
specific \gls{CPS} is needed to attack it; however, the attacker still
requires the minimal knowledge of the \gls{CPS} itself as well as the reactive
power injection control rule, i.e., minimal domain-specific knowledge. The
concept of \gls{ARL} by \textcite{Fischer2019arl} goes beyond specific \gls{CPS}
domains in that it does not require domain-specific knowledge, but only a
minimal description of sensors and actuators of an agent, where \emph{minimal}
means the mathematical description of the observation or action space,
respectively, such as ``discrete \(1, 2, \dotsc, 10\)'' or ``continuous \([0;
1]\)''. In \gls{ARL}, attacker and defender agents work against each
other---but without knowledge of their respective counterpart---on
a shared model of a \gls{CPS}, training each other in the process. The
attacker tries to de-stabilize the \gls{CPS} (where the notion of stability is
up to the experimenter to define), whereas the defender tries to ensure a
resilient operation of the \gls{CPS}. Over time, the attacker finds attack
vectors, whereas the defender learns strategies to keep the \gls{CPS}
operational. This is the second distinguishing feature of \gls{ARL}, in that
it provides not only analysis through the attacker, but also tries to produce
a operation strategy.

\section{Conclusion}
\label{sec:conclusion}

In this paper, we have provided a literature review of techniques for
\gls{CPS} design and analysis, beginning with the basics of temporal logic and
requirements that serve to formally specify the \gls{CPS}' components at hand.
We have touched methods for program synthesis, i.e., ways to generate programs
from a formal specification. We have then extensively covered the use of
\glspl{ANN} as controllers, e.g., as replacements for classical control loops,
and how falsification is applied to cover the safety requirements of a
\gls{CPS}. We have seen that \gls{DL} has great potential, but also increases
the uncertainty of such systems considerably. Going from component to system
level, we have covered simulation frameworks for \glspl{CPS}. Afterwards, we
have broadened the perspective, and included \glspl{MAS} as communicating,
distributed problem-solvers that nowadays not only can, but do control vast
\glspl{CPS}. Finally, we have considered attack vectors in terms of malicious
actions against a \gls{CPS}.

Obviously, \gls{AI} research in perspective of \glspl{CPS} introduces much
uncertainty and often breaks the required rigid assertion of liveness and,
especially, safety requirements in this domain. The benefits make the risk
beneficial, can by no means eschew the need for developing assertion methods
towards a verified \gls{AI}. We also believe that reversing the situation and
putting \gls{AI} to use to analyse a \gls{CPS} for its safety requirements,
even without \gls{CPS} domain knowledge required, will provide a valuable path
for a beneficial collaboration of the two research domains.

\section{Acknowledgements}

The authors would like to thank Sebastian Lehnhoff for his counsel, strategic
advice and support of the \gls{ARL} project, and Eckard Böde for reviewing
initial drafts of this paper, providing valuable inputs.

\printbibliography
\end{document}